

\def\figflag{I}

\input harvmac.tex
\def\figI{I}
\ifx\figflag\figI \input epsf.tex \fi

\Title{\vbox{\baselineskip12pt
\hbox to \hsize{\hfill NSF-ITP-92-155}
\hbox to \hsize{\hfill UCSBTH-92-47}
\hbox to \hsize{\hfill hep-th/9211118}}}
{\vbox{\centerline{Low-Energy Dynamics of String Solitons}}}
\centerline{Andrew G. Felce\footnote{$^\dagger$}
{felce@nsfitp.itp.ucsb.edu}}
\medskip
\centerline{\it Institute for Theoretical Physics}
\centerline{\it University of California, Santa Barbara, CA 93106}
\bigskip
\centerline{T. M. Samols}
\medskip
\centerline{\it Department of Physics, University of California}
\centerline{\it Santa Barbara, CA 93106}
\vskip .5in
\centerline{\bf Abstract}
\smallskip
The dynamics of a class of fivebrane string solitons is considered in
the moduli space approximation.  The metric on moduli space is found to
be flat. This implies that at low energies the solitons do not
interact, and their scattering is trivial.  The range of validity
of the approximation is also briefly discussed.
\Date{11/92}
\vfill\eject

\def\gd{\dot\gamma}
\def\gdij{\dot g_{ij}}
\def\gdkl{\dot g_{kl}}

\def\phid{\dot\phi}
\def\Bdij{\dot B_{ij}}
\def\Bdkl{\dot B_{kl}}

\def\vd{\dot\varphi}

\def\quart{{1\over4}}
\def\sqg{\sqrt{g}~}
\def\sqh{\sqrt{h}~}
\def\dij{\delta_{ij}}
\def\gij{g_{ij}}

\def\Bmn{B_{\mu\nu}}
\def\apm{\alpha^{\prime}}
\def\ka{\kappa}
\def\la{\lambda}
\def\e{\epsilon}
\def\xui{\dot\xi^i}
\def\xuj{\dot\xi^j}
\def\xuk{\dot\xi^k}
\def\xdi{\dot\xi_i}
\def\xdj{\dot\xi_j}

\def\xdm{\dot\xi_m}
\def\deldi{\nabla_i}
\def\deldj{\nabla_j}
\def\deldk{\nabla_k}

\def\delui{\nabla^i}
\def\deluj{\nabla^j}

\def\delul{\nabla^l}
\def\tdeldi{\tilde\nabla_i}
\def\tdeluj{\tilde\nabla^j}
\def\coeff{{2\pi^2 \over\ka^2}}
\def\wdijkl{\omega_{ijkl}}

\def\MN{{\cal M}_N}
\def\an{\vec{a}_n}

\lref\neut{A. Strominger, Nucl. Phys. {\bf B274} (1986) 253.}
\lref\dufflu{M. J. Duff and J. X. Lu, Nucl. Phys. {\bf B354} (1991) 141.}
\lref\CHS{C. G. Callan, J. Harvey and A. Strominger,
Nucl. Phys. {\bf B359} (1991) 611.}
\lref\witt{R. Rohm and E. Witten, Ann. Phys. {\bf 170} (1986) 454.}
\lref\ramzi{R. R. Khuri, Nucl. Phys. {\bf B376} (1992) 350.}
\lref\callram{C. G. Callan and R. R. Khuri, Phys. Lett. {\bf 261B}
(1991) 363.}
\lref\york{J. W. York, Phys. Rev. Lett. {\bf 28} (1972) 1082.}
\lref\ruback{P. J. Ruback, Comm. Math. Phys. {\bf 107} (1986) 93.}
\lref\ADM{R. Arnowitt, S. Deser and C. W. Misner,
in {\it Gravitation: An Introduction to Current Research},
ed. L.~Witten (Wiley, New York, 1962).}
\lref\mant{N. S. Manton, Phys. Lett. {\bf 110B} (1982) 54.}
\lref\priv{A. Strominger, private communication.}
\lref\strom{A. Strominger, Nucl. Phys. {\bf B343} (1990) 167.}
\lref\hhs{J. H. Horne, G. T. Horowitz and A. R. Steif,
Phys. Rev. Lett. {\bf 68} (1992) 568.}
\lref\nestberg{J. Isenberg and J. Nester, in Vol. 1 of {\it General
Relativity and Gravitation}, ed. A.~Held (Plenum, New York, 1980).}

This note is concerned with the low energy dynamics of string theory
solitons -- more precisely the neutral wormhole fivebranes~\refs{
\neut, \dufflu, \CHS}.  The method, which has now been
applied to a number of soliton systems, is to approximate
the field evolution as geodesic motion on the moduli space of
static solutions, the metric being that induced by the
field kinetic energy~\mant. We find that the metric
is flat, implying trivial scattering at low energies.
This result was conjectured in~\callram, and is in agreement
with more general considerations we shall mention later.
The same problem has also been addressed in~\ramzi, where however
a different result was obtained.

We first briefly review the soliton solutions of interest.  The bosonic
part of the low-energy effective action for Type-II string theory
(or heterotic strings when the gauge field $F$ is zero) in ten dimensions
is given to lowest order in the Regge slope $\apm$ by
\eqn\lowen{I={1\over2\ka^2}\int d^{10}x\sqg e^{-2\phi}
                \bigl[R+4(\nabla\phi)^2-{1\over3}H^2\bigr]
		~~+~~{\rm Surface~Term}~,}
where $\ka^2$ is proportional to the gravitational constant,
$\phi$ is the dilaton field, and $H_{\la\mu\nu}$ is the
field strength obtained from the antisymmetric tensor field $\Bmn$
($H_{\la\mu\nu}=\half\del_\la \Bmn + {\rm cyclic})$.
The corresponding equations of motion are
\eqn\grav{\eqalign{R_{\mu\nu}+2\nabla_\mu\nabla_\nu\phi -
                                   H_{\mu\nu}^2 &=0~,\cr
   \nabla^{\la}\bigl(e^{-2\phi}H_{\la\mu\nu}\bigr) &= 0~, \cr
   \nabla^2\phi - 2(\nabla\phi)^2 + {1\over3}H^2 &=0~.}}
Restricting immediately to fields with five-dimensional translational
symmetry, the system is essentially (4+1)-dimensional, and the
remaining dimensions may be ignored.  Field configurations are then
characterized by the total flux $Q$ (the ``axion charge''), of the
$H$-field through the 3-sphere at infinity.  For a given $Q$, the ADM
mass\foot{Without loss of generality we set the asymptotic
dilaton field to zero. The ADM mass is then the same in both
canonical and sigma-model variables.} satisfies the
Bogomol'nyi-type bound~\strom\
\eqn\Bog{M_{ADM} \ge \coeff Q~.}
Saturating this bound gives the neutral fivebrane soliton solutions.
Up to coordinate and gauge transformations, the general solution is
\eqn\met{\eqalign{g_{00}&=-1\qquad\qquad\qquad\qquad\gij=e^{2\phi}\dij~,\cr
    e^{2\phi} &= 1 + \sum_{n=1}^N {Q_n\over|\vec{x}-\vec{a}_n|^2}
			\qquad\qquad
			H_{ijk}=\wdijkl\delul\phi~,}}
all other components being zero. Here, $i$ runs from $1$ to $4$,
$\vec{x}$ is the position vector in 4-dimensional space,
and $\wdijkl$ is the volume form on this space.  Physically, \met\
describes a static configuration of $N$ solitons with positions $\an$
and charges $Q_n$.  Each soliton is a semi-infinite wormhole,
stabilized by the flux of the $H$-field through its throat (see
Fig.1); one may regard the existence of the multi-soliton static
solution as a consequence of the exact cancellation between the
attractive forces due to gravity and the dilaton, and a repulsive
force due to the $H$-field.  The $Q_n$'s are quantized in units of
$\apm$~\witt, and so for given values, the moduli space of solutions,
$\MN$, is just the space of $\an$'s -- in other words, ${\bf R}^{4N}$.

{\midinsert
\vbox to 2.2truein{\ifx\figflag\figI
\vfil
\centerline{\epsfxsize=3.0in \epsfbox{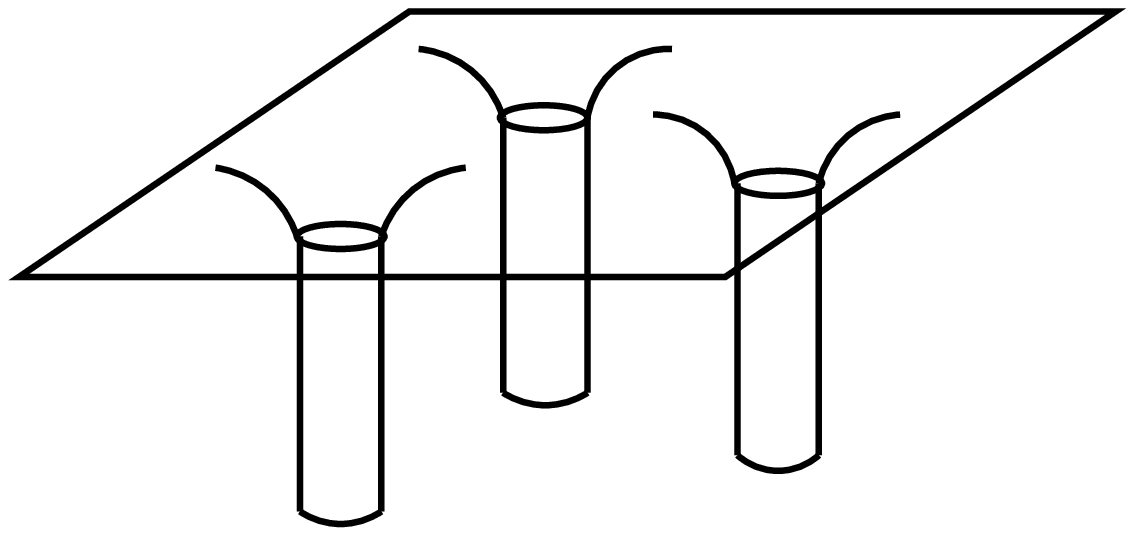}}
\vfil\fi}
\baselineskip=12pt
\centerline{\hbox{\vbox{
\hbox{\bf Fig. 1: }
\hbox{}}
\vbox{
\hbox{Two-dimensional cross-section of a typical 3-soliton}
\hbox{configuration.  (The wormhole throats are infinitely long.)}}}}
\endinsert}

To provide the setting for the moduli space description of the dynamics
we perform a split between space and time, extending a treatment given
by Ruback for a similar analysis of Kaluza-Klein monopoles~\ruback.  We
assume that the low energy dynamics involves only small perturbations
away from $\MN$.  In the neighbourhood of a space-like surface $\Sigma$
we may choose synchronous coordinates, so that $g_{00}=-1$, $g_{0i}=0$,
and the metric takes the form\foot{In fact, under small perturbations,
such as those caused by imparting small velocities to the solitons,
event horizons will form at some distance down the wormhole throats.
Our assumption is that these horizons are sufficiently far down that
they cause no significant effects.}
\eqn\synch{ds^2=-dt^2 + g_{ij}dx^i dx^j~.}
The $U(1)$ gauge-invariance associated with the antisymmetric tensor
field allows in addition the choice $B_{0i}=0$.  Performing standard
manipulations~\ADM, the action may be written as the time integral of
the Lagrangian $L=T-V$ where
\eqn\Vdef{\eqalign{V = -{1\over2\ka^2}\int\limits_\Sigma d^4x\sqg
		e^{-2\phi}\bigl[^4R + &4\delui\phi\deldi\phi
		- {1\over3}H^{ijk}H_{ijk}\bigr]\cr
		&+{1\over\ka^2}\int\limits_{\partial\Sigma} d^3x\sqh
		e^{-2\phi}(K^S-K^S_0)}}
is to be regarded as the potential energy, and
\eqn\Tdef{T = {1\over8\ka^2}\int\limits_\Sigma d^4x\sqg e^{-2\phi}
		\bigl[g^{ik}g^{jl}(\gdij+\Bdij)(\gdkl+\Bdkl)
		-(\gd-4\phid)^2\bigr]}
the kinetic energy.  In these expressions, $^4R$ is the
four-dimensional Ricci scalar evaluated on the spatial metric, $K^S$
is the extrinsic curvature scalar of the boundary at spatial infinity,
$h$ is the induced metric on the boundary, and $K^S_0$ is the curvature
scalar that the boundary would have were it embedded in flat space.
Dots denote time derivatives, and $\gd=g^{ij}\gdij$. The surface term
in~\Vdef\ is a remnant of the surface term in~\lowen, which is
required to compensate for the presence of second derivatives in the
Ricci scalar~\york; the surface terms involving time derivatives
cancel against a total time derivative present in the expression
relating the five-dimensional Ricci scalar to $^4R$.

We must also take care to impose the equations of motion corresponding
to $g_{00}$, $g_{0i}$ and $B_{0i}$.  These are just the constraints
associated with the diffeomorphism invariance and $U(1)$ gauge
invariance of the theory.  The Hamiltonian and momentum constraints
associated with the former are
\eqn\Hamcon{\eqalign{\quart g^{ik}g^{jl}(\gdij+\Bdij)&(\gdkl+\Bdkl)
		-\quart(\gd-4\phid)^2 \cr& - ^4R - 4\delui\deldi\phi +
             4\delui\phi\deldi\phi + {1\over3}H^{ijk}H_{ijk}=0}}
and
\eqn\gravcon{\tdeluj\bigl[e^{-2\phi}(\gdij+\Bdij)\bigr]=
		e^{-2\phi}\tdeldi(\gd-4\phid)}
respectively, where $\tdeldi$ is a generalized covariant derivative in
which the Christoffel connection $\Gamma^i_{~jk}$ is replaced by
$\tilde\Gamma^i_{~jk}=\Gamma^i_{~jk}-H^i_{~jk}$; the (Gauss) constraint
associated with the latter is
\eqn\Hcon{\deluj\bigl[e^{-2\phi}\Bdij\bigr]=0~.}

Setting aside the Hamiltonian constraint for a moment,
it is useful to interpret these expressions geometrically.
We regard the kinetic energy~\Tdef\  as defining a
metric on the space of fields $f=(g_{ij},\phi,B_{ij})$.
In this space one must of
course regard as equivalent the points along each orbit
generated by diffeomorphisms and $U(1)$ transformations.
The momentum and Gauss constraints, \gravcon\ \Hcon, restrict
tangent vectors, $\dot f$, to be orthogonal
to these orbits with respect to the metric defined by $T$.
However, this still leaves some gauge freedom in $\dot f$.
In particular we can add a vector corresponding to an infinitesimal
diffeomorphism generated by a vector field (on $\Sigma$) of the
form $\dot\xi_i=\nabla_i\lambda$.  By an appropriate choice
of $\lambda$ we may set
\eqn\traceless{\gd - 4\phid = 0~.}
This condition fixes the gauge completely, and renders the kinetic
energy $T$ positive definite. In summary then, we have a gauge invariant
positive definite metric, and potential $V$.
Furthermore, if the Hamiltonian constraint \Hamcon\ is satisfied,
$T+V$ is given by a boundary term which can be shown to be equivalent to
the ADM mass $M_{ADM}$.\foot{The boundary term consists of the boundary
part of~\Vdef, together with a piece which involves the normal derivative
of the dilaton.  The former is equivalent to the canonical ADM
formula~\nestberg; the latter is precisely the term which generalizes
this for the sigma-model metric~\hhs.}

The idea of the moduli space description is that given
initial data corresponding to giving the solitons some
small velocities ({\it i.e.} a slow motion tangent to $\MN$),
$V$ will force the motion to remain close to $\MN$, and the evolution
will be well approximated by geodesic motion on $\MN$ with respect to
the metric induced by $T$. This requires of course that $\MN$ is at
a (local) minimum of $V$. To see that this is indeed the case, consider
a small perturbation away from a point $f_0$ of the moduli space:
\eqn\pert{f=f_0+\e v + O(\e^2) \qquad \dot f =0~,}
where the tangent vector $v$ satisfies \gravcon, \Hcon\ and \traceless.
To $O(\e)$, the Hamiltonian constraint~\Hamcon\ is simply the
divergence of the momentum constraint~\gravcon\ evaluated on $v$ and is
thus automatically satisfied.  The potential $V$ is then just the ADM
mass, and the Bogomol'nyi bound~\Bog\ gives the result.

To calculate the metric on $\MN$ we require an expression for a general
tangent to $\MN$ satisfying the constraints as well as~\traceless.  Such
a vector is given by
\eqn\metvar{\gdij=2\gij\vd+\deldi\xdj+\deldj\xdi~,}
\eqn\dilvar{\phid=\vd+\xuk\deldk\phi~,}
and
\eqn\Bvar{\Bdij=\omega_{ij}^{~~km}\deldk\xdm+2\xuk H_{ijk}~,}
together with~\met, where
\eqn\vphi{\vd=\sum\limits_{n=1}^N {\del\phi\over\del a^i_n} \dot a^i_n~,}
and
\eqn\gauge{\xdi=\sum_{n=1}^N{Q_n \dot a_n^i
		\over|\vec{x}-\vec{a}_n(t)|^2}~.}
The first term in each case represents the variation of the
fields~\met\ by an infinitesimal change in the moduli; the other terms
represent the effect of an infinitesimal diffeomorphism generated by
$\xuk$ together with an infinitesimal $U(1)$ rotation.
This is just the solution obtained in~\ramzi\ by giving each
soliton an independent boost, but in a different gauge.
(An appealing feature of this gauge is that the fractional variations of
the fields are small at every point in space, in particular near the
soliton cores.  This makes regularization unnecessary.)
It is straightforward to check the required properties.  The
condition~\traceless\ is equivalent to the identity
\eqn\vddef{\vd=-\half\deldi\xui + \xui\deldi\phi~.}
Furthermore, since we are on $\MN$, the Hamiltonian constraint is
automatically satisfied to $O(\dot{\vec{a}}_n)$, while
the Gauss constraint is automatically satisfied by~\Bvar\ for any
$\xui$.  The momentum constraint~\gravcon\ is a little more tricky,
but can be reduced to the form
\eqn\soltest{\half\deluj\bigl(\deldj\xdi-\deldi\xdj\bigr)
		= e^{-2\phi}\deldi\bigl(e^{2\phi}\vd\bigr)~.}
Expressed in terms of ordinary derivatives, this is equivalent to
\eqn\flatcon{e^{-2\phi}\bigl[\del^2\xdi-\del_i
			(\del_j\xdj+2e^{2\phi}\vd)\bigr]=0}
where $\del^2$ is the flat space Laplacian, and it is straightforward to
check that $e^{-2\phi}\del^2\xdi$ and $\del_j\xdj+2e^{2\phi}\vd$
vanish separately.

All that now remains is to substitute~\metvar\
--~\Bvar\ into the expression~\Tdef\ for the kinetic energy.
Using~\vddef, the kinetic energy integrand can be written
\eqn\calt{\eqalign{\delui\bigl[&e^{-2\phi}\xuj\deldi\xdj\bigr]\cr
			-\xui &e^{-2\phi}\bigl[
			\nabla^2\xdi-\deldj\deldi\xuj-2\deldi\vd-
			4\vd\deldi\phi\bigr]\cr
			-\xui\xuj &e^{-2\phi}\bigl[
			R_{ij}+2\deldi\deldj\phi-H_{ij}^2\bigr]~.}}
The second term vanishes by \soltest. The third term is just the
graviton equation of motion for the static solution~\grav\ and is also
zero. The final answer is thus given by a surface integral
\eqn\Tans{T={1\over4\ka^2}\int\limits_{\partial\Sigma}d^3x\sqh e^{-2\phi}
		\nabla_{\bf n} \dot\xi^2,}
where $\nabla_{\bf n}$ is the normal derivative on the surface.
Examination of the integrand reveals that the surface at space-like
infinity contributes nothing, but that there is a finite contribution
from the asymptotic $S^3$ associated with each wormhole throat.  A simple
calculation shows that the result is
\eqn\kin{T=\sum_{n=1}^N\half M_n |\dot{\an}|^2}
where $M_n=\coeff Q_n$ is the ADM mass of a single soliton of charge
$Q_n$.  This means that the total kinetic energy is just the sum of the
individual energies with no interaction terms.  The metric on moduli
space is therefore flat, and the low energy scattering trivial.\foot{
The result is in agreement with more general considerations~\priv.
The Type-II soliton we have been considering possesses $N=4$ worldsheet
supersymmetry~\CHS.  It can be argued that this implies that the moduli
space is hyperK\"ahler and so Ricci flat.  The moduli space for two
solitons (factoring out the center of mass motion) is topologically
${\bf R}^4$, and for this case at least Ricci flat is equivalent to flat.}

It remains to consider the validity of the approximation. For
general scattering processes one expects the main corrections
to be due to radiation in both the gravitational and matter
fields, with scalar radiation dominating. In very close collisions
however, the approximation will break down more seriously.
Consider two solitons in collision. Since their total energy
exceeds the Bogomol'nyi bound by the kinetic energy of the
relative motion, there will be an event horizon for the
combined system.  For initial data with some non-zero impact
parameter there will be a critical velocity below which the treatment
we have given is valid, but for velocities which exceed this value,
the solitons will fall within the horizon, and will therefore coalesce.

\vfill\eject
\vskip .5in
\centerline{\bf Acknowledgements}
\bigskip
We thank C. Callan, G. Horowitz, A. Strominger and R. Tate for useful
discussions.  This work was supported by the National Science Foundation
under grants PHY89-04035 and PHY91-57463, and by the Department of
Energy under grant 91ER40618.


\vskip .5in
\footatend\vfill\immediate\closeout\rfile\writestoppt
\baselineskip=14pt\centerline{{\bf References}}\bigskip{\frenchspacing%
\parindent=20pt\escapechar=` \input \jobname.refs\vfill\eject}\nonfrenchspacing

\bye